# Penetration Deep into Tissues of Reactive Oxygen Species Generated in Floating-Electrode Dielectric Barrier Discharge (FE-DBD): *in Vitro* Agarose Gel Model Mimicking an Open Wound

Danil Dobrynin[1], Gregory Fridman[2,1], Gary Friedman[3,1] and Alexander Fridman[4,1]

[1] *A.J. Drexel Plasma Institute, Drexel University*

[2] *School of Biomedical Engineering, Drexel University*

[3] Electrical and Computer Engineering Department, *College of Engineering, Drexel University*

[4] *Mechanical Engineering and Mechanics Department, College of Engineering, Drexel University*

**Abstract**: In this manuscript we present an *in vitro* model based on agarose gel that can be used to simulate a dirty, oily, bloody, and morphologically complex surface of, for example, an open wound. We show this model's effectiveness in simulating depth of penetration of reactive species generated in plasma (e.g. $H_2O_2$) deep into tissue of a rat and confirm the penetration depths with agarose gel model. We envision that in the future such a model could be used to study plasma discharges (and other modalities) and minimize the use of live animals: plasma can be optimized on the agarose gel wound model and then finally verified using an actual wound.

## Introduction

Plasma Medicine is now a rapidly advancing field with positive indication of plasma's ability to sterilize many types of surfaces [1-4], including human and animal tissue [5], treat wounds [6, 7], coagulate blood [5, 8], and even treat diseases [7]. This is mostly likely due to the plasma-catalysis of the reduction/oxidation reactions happening in the biological system. Action of specific charged or neutral active species or radiation is frequently associated with the corresponding specific effect (e.g., anti-inflammatory effect of nitric oxide (NO), and highly oxidative hydroxyl radicals and other reactive oxygen species (ROS)). With the vast amount of recent research it is becoming clear that plasmas are, indeed, able to induce some, perhaps positive, clinical effects in patients. Thus, it is important to analyze plasma itself, not only its effects on the organism. Clearly, plasma in contact with dirty, oily, bloody tissue is not the same as the plasma generated between two perfectly controlled electrodes. For this reason we have developed a simple *in-vitro* model where the second electrode for plasma generation is an agarose gel. In this manuscript we compare measurements of plasma-generated species on agarose gel to those on store-bought chicken meat and to those on a live rat wound. Surprisingly, the findings reported here suggest that properly-prepared agarose gel may actually serve as a very good model of penetration of plasma-generated species into the tissue. Indeed, if our findings are true and are confirmed by other research groups, we may begin to analyze plasma (perform spectroscopic, microwave, and other measurements) with the second electrode being a simple agarose gel model, with these measurements being a sufficiently accurate representation of the real environment. We leave it to the reader to judge our findings and urge them to verify our results.

## Materials and methods

In this study we used atmospheric pressure dielectric barrier discharge (DBD) plasma at room temperature in air. DBD plasma was generated using an experimental setup similar to the one previously described elsewhere [9-13]. In short, the discharge was generated by applying alternating polarity pulsed (1 kHz) voltage of ~20 kV magnitude (peak to peak) and a rise time of 5 V/ns between the insulated high voltage electrode and the sample undergoing treatment. The powered electrode was made of a 1.5 cm diameter solid copper disc covered by a 1.9 cm diameter 1 mm thick quartz dielectric. The discharge gap was kept at 1.5 mm. Current peak duration was 1.2 μs, and corresponding plasma surface power density was 0.3 W/cm$^2$. In the case of *ex vivo* measurement in a rat tissue, a special pen-size electrode was used: one millimeter thick polished clear fused quartz (Technical Glass Products, Painesville, OH) was used as an insulating dielectric barrier, where a handheld pen-like device with the quartz tip was used for treatment. In this case, the average power density for the active area of the high voltage electrode was kept at the level of approximately 0.74 Watt for 6 mm electrode diameter.

Agarose gels were prepared using standard procedure with pure agar powder (Fisher) in either distilled water or phosphate buffered saline (PBS, Fisher). In order to determine the best concentration of agarose gels which would closely represent tissue, we have used agar at concentrations of 0.6%, 1.5% and 3% weight percentage. These values were chosen for the following reasons. Agarose gels at 0.6% concentration were reported to closely resemble *in vivo* brain tissue with respect to several physical characteristics [14]. Four percent agar phantoms are widely used as a tissue models for radiology studies [15]. The 1.5% concentration of agar was chosen as a median point which is often used as a microbiological substrate.

Measurements of $H_2O_2$ and pH penetration into agarose gels (0.6%, 1.5%, and 3% wt) and tissues were done using Amplex UltraRed reagent (Invitrogen, ex/em: 530/590 nm) and Fluorescein (Sigma Aldrich, ex/em: 490/514 nm) fluorescent dyes respectively. In the case of $H_2O_2$, 75μL PBS containing 100μM Amplex UltraRed with 200 U/μL horseradish peroxidase (MP Biomedicals) were placed in between 1 mm thick 4x4 cm agar slices and incubated for about 15 minutes before the treatment in order to provide presence of the dye in the agar volume; for the pH measurement, the agarose gels were prepared by adding fluorescein dye before its solidifying. In order to measure the $H_2O_2$ and pH in tissue, the dyes were injected using a syringe into a 1 cm thick 4x4 cm skinless chicken breast tissue sample at various points to the depth of up to 1 cm. *Ex vivo* measurements were done in a rat tissue: animal (hairless Sprague-Dawley male rat) was euthanized right before the procedure. 200 µl of dye solution (Amplex UltraRed) was injected subcutaneously using a sterile syringe, and the animal skin was treated with FE-DBD plasma for various time points after 5 minutes incubation period (Figure 1). Right after the treatment, skin tissue samples were extracted and analyzed as follows. Treated samples were sliced in a vertical direction with thickness of 1 mm, and fluorescence was measured using an LS55 (Perkin Elmer) fluorescent spectrometer equipped with XY reader accessory (Figure 2-Figure 3). To obtain calibration curves for hydrogen peroxide in the plasma treated samples a properly diluted standard stabilized 3% $H_2O_2$ (Fisher) water solution was used to obtain various concentrations.

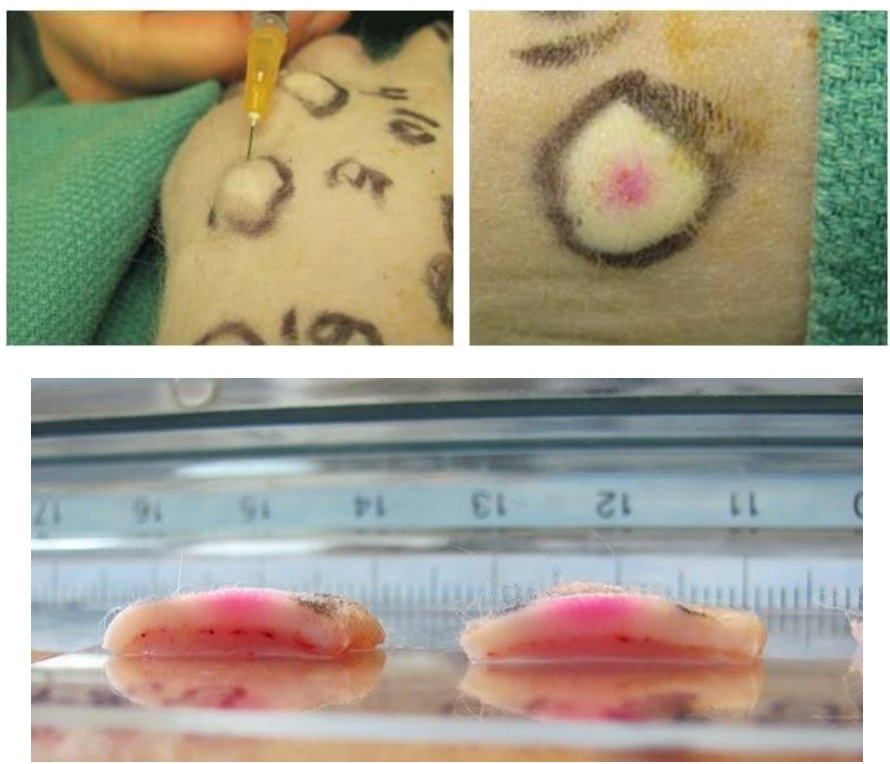

Figure 1. Subcutaneous injection of the fluorescent dye (euthanized rat, top left), the rat skin after the plasma treatment (top right), and the skin sample cross-section just before measurement (bottom).

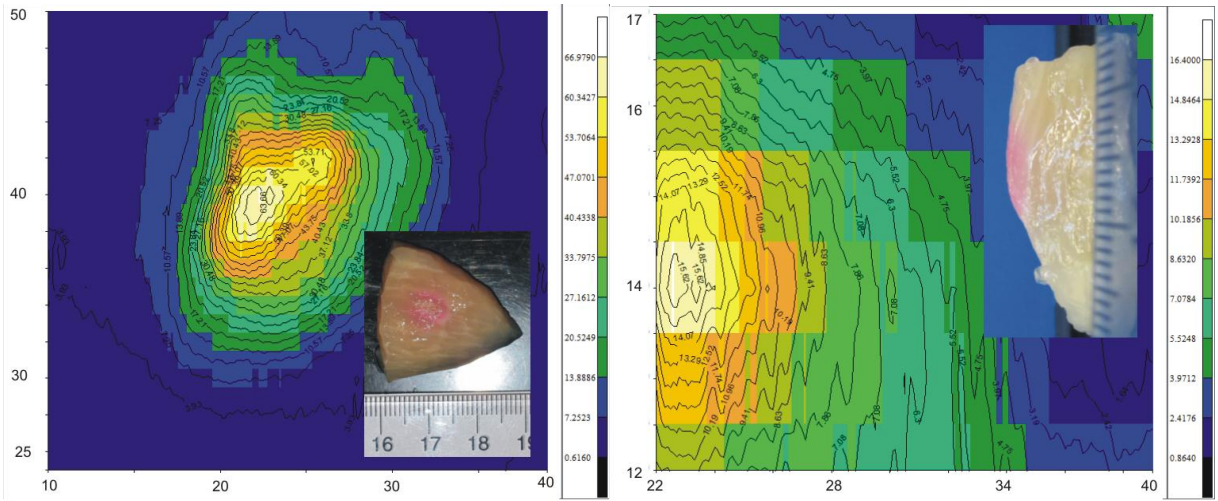

Figure 2 Chicken breast after plasma treatment with $H_2O_2$ fluorescent dye: photograph and fluorescent images from the top (left) and side (right) of the sample (color intensity in arbitrary units, dimensions in mm).

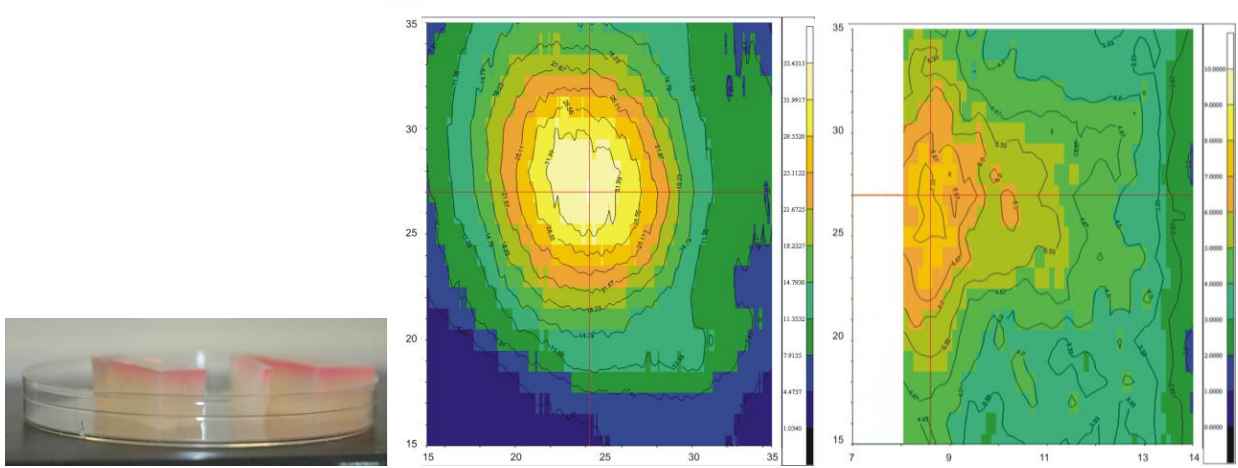

Figure 3 Agarose gel after plasma treatment with $H_2O_2$ fluorescent dye: photograph and fluorescent images from the top and side of the sample (arbitrary units, dimensions in mm).

## Results

The results of $H_2O_2$ measurements in dead tissue are shown on Figure 4: with longer treatment time depth of penetration as well as increasing concentrations of hydrogen peroxide. In general, several mM of $H_2O_2$ are produced in tissue after plasma treatment, while it diffuses 1.5-3.5 mm deep. Roughly the same tendency is observed in the case of tissue acidity change, shown on Figure 5 (florescence intensity of fluorescein decreases with lower pH, and the data are presented in arbitrary units); however, the effect of pH lowering penetrates deeper – up to 4.5-5 mm.

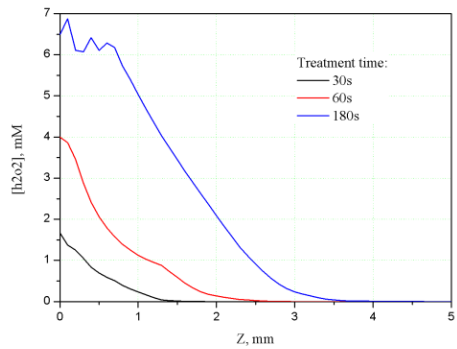

Figure 4 The profiles of $H_2O_2$ in tissue after the plasma treatment.

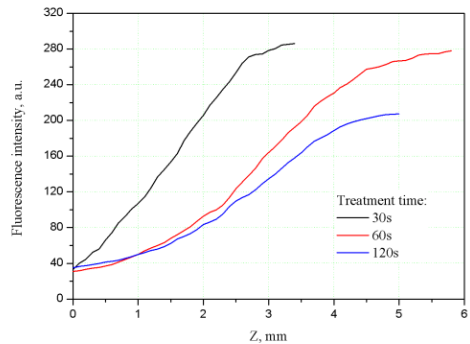

Figure 5 The profiles of pH in dead tissue after the plasma treatment.

In order to develop a simple realistic in vitro model of tissue that would have simple physicochemical characteristics, primarily from the point of view of depth of reactive species penetration, three concentrations of agarose media were used. The measurement results for $H_2O_2$ produced by plasma treatment in agar gels together with the results for dead tissues for the same treatment doses are shown on Figure 6. Hydrogen peroxide concentration on the agar gel surface varied with different agar densities: 0.5 mM for 0.6%, 0.7 mM for 3% and 1.9 mM for 1.5% gels after 1 minute of plasma treatment. The results of hydrogen peroxide penetration measurements in euthanized rat skin tissues are shown on Figure 7, and compared to the dead chicken breast tissue, depth of penetration appears to be very similar – up to 4 mm after 2 minute treatment (although concentration is almost an order of magnitude greater, due to the power of the discharge being 8 times higher). Surprisingly, depth of $H_2O_2$ penetration for all types of agarose media was about the same: Figure 8 shows depths at which the same level of 0.05 mM of $H_2O_2$ was detected in agar gels and tissue for different plasma exposure times.

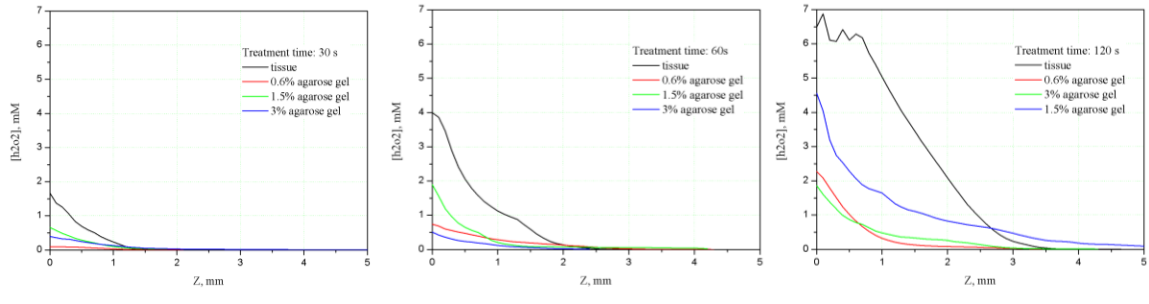

Figure 6 The profiles of $H_2O_2$ in non-buffered agarose gels and dead tissues after the plasma treatment.

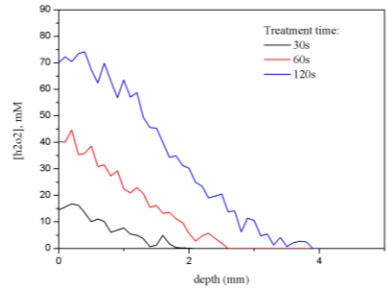

Figure 7 The profiles of H2O2 in euthanized rat tissue after the plasma treatment.

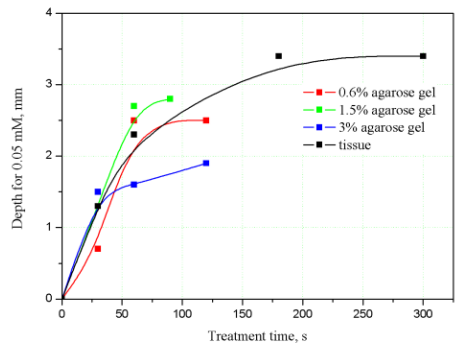

Figure 8 Depth of $H_2O_2$ penetration at concentration of 0.05 mM: comparison for agarose gels and dead tissue

In contrast to hydrogen peroxide, the dynamics of acidity change inside of agar gels fluctuate significantly with each agar media composition (Figure 9). Tissue appeared to have better buffer properties compared to non-buffered agarose gels. However, addition of PBS into agar displays expected similarities in terms of pH changes.

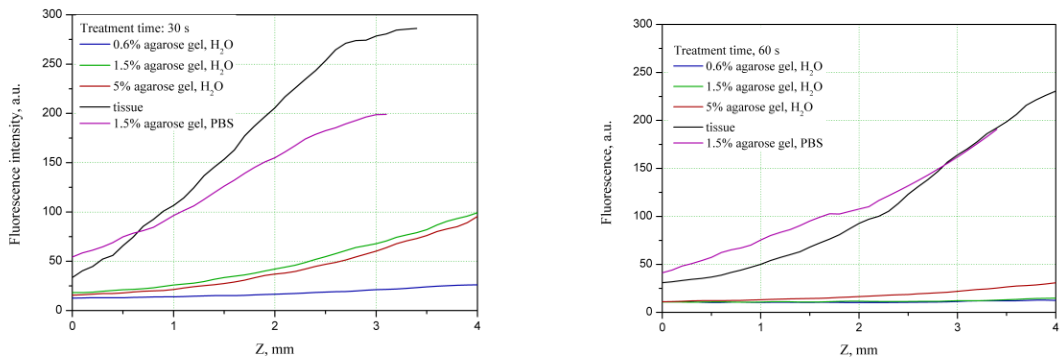

Figure 9 The profiles of pH in agarose gels and dead tissues after the plasma treatment.

## Discussion

In this paper we address a problem of direct atmospheric pressure plasma interaction with tissues; particularly, delivery of neutral active components produced by plasma inside of tissues, and creation of a physical model of such interaction. Plasma is a complex chemically active medium, and a number of biomedical studies suggest that it can be successfully applied to various living tissues without damage [16-18], even providing a number of positive effects, such as blood coagulation [10, 13, 19, 20], wound [13, 19, 21, 22] and even cancer treatment [12, 23, 24], for example. Notably, for instance in the case of experimental *subcutaneous* cancer tumor treatment in a mouse model [24], the effect of plasma treatment is observed not only on the surface of treated tissue (skin), but also (and chiefly) in volume, under the skin. In many cases, researchers report a significant change of pH of solutions and tissues after plasma treatment (see, for example, [24, 25]). On the other hand, hydrogen peroxide, which is almost always produced by plasma, is often considered to be important in bacteria inactivation [9, 25-28] or wound healing processes [29]. Therefore, in the current study we focused on measurements of these two parameters – pH and $H_2O_2$ concentration - as a function of treatment time and depth of penetration into a treated object.

The evaluation of measuring techniques on the effects of plasma treatment on tissues is complicated in direct experiments on actual biological objects due to a wide variation of morphological and biochemical parameters that are beyond the control of the experimenter. Therefore, stable and reproducible test objects, that mimic tissue physicochemical characteristics, are needed. In the current study, we have used a custom-made agarose gel, similar to those used in microbiological studies, but modified to achieve certain similarities with dead tissue. Agarose phantoms are widely used as models of various tissues (see for example, [14, 15, 30, 31].

The results we present in this paper show that, in the case of real tissue, active species produced by plasma on the surface may travel in tissue volume to a depth of several millimeters. Obviously, this depth of penetration is determined by diffusion and reaction rates, which would highly depend on type of tissue and its biochemical characteristics. A measurement of these parameters is an extremely complicated task; and therefore, creation of a simple model that would closely represent certain tissue is possible experimentally. We used agarose gels of various concentrations in order to mimic the physical properties of tissue. For the case of hydrogen peroxide, as shown in Figure 6 and Figure 8, depth of diffusion is about the same for all types of agar, but the concentration of $H_2O_2$ varies, being the closest for the 1.5% agar weight percent case, and about 3 times lower for both 0.6% and 3% wt agar gels. This behavior may possibly be explained by both diffusion properties and reaction rates of $H_2O_2$ in agar.

Acidity of tissue (in contrast to the experimental *in vivo* data reported in [24] where authors observed a significant decrease of pH on the skin surface with a slight increase in subcutaneous tissue), in our experiments was observed to consistently increase (lowering of pH) as a result of exposure to the discharge. Since the fluorescence of fluorescein highly depends not only on pH of a solution, but also on other parameters [32-36], we did not attempt to obtain calibration curves; and therefore the data are presented in arbitrary units. We show that tissue compared to agarose gels prepared in distilled water acts as a significantly better buffer, and that the depth of pH changes inside such gels is much greater. In fact, we observe a significant drop in the whole volume of a phantom – up to a centimeter thick agar. This problem, as shown on Figure 9, may be addressed simply by adding a buffer into the agarose media – 1.5% wt agar was prepared in 1X PBS.

In summary, we show that plasma effects may be transferred several millimeters deep inside a tissue, as measured in an *ex vivo* chicken tissue and rat skin models. We have detected a penetration behavior of two simple active components, namely hydrogen peroxide and pH, but other species may be detected and measured using other fluorescent dyes or techniques. In addition, we have shown that a simple agar gel model may express similar physicochemical properties as a real tissue, resulting in comparable penetration effects of active species.

## Conclusion

In this study we utilized atmospheric pressure dielectric barrier discharge treatment of living rat tissue and we have developed a corresponding agarose gel model that mimics the tissue. Our findings suggest that with careful preparation of the agarose gel it may serve as a very good model of penetration of plasma-generated active species (such as reactive oxygen and reactive nitrogen species) into the tissue. We show that this simple model can be accurately used now for further plasma studies (spectroscopic measurements, microwave diagnostics, etc) without the need to use a live animal for each experiment. Potentially, this may significantly reduce the regulatory hurdles in performing plasma medical experiments, reduce the number of animals needed to complete the study, and ultimately generate a larger bank of experimental data. The data reported in this manuscript on penetration on

$H_2O_2$ into the animal tissue can be reproduced with custom-made agarose gel. These findings need to be validated for other plasma-generated active species as well and this work is currently underway.

# References

1. Cooper, M., G. Fridman, D. Staack, A.F. Gutsol, V.N. Vasilets, S. Anandan, Y.I. Cho, A. Fridman, and A. Tsapin, *Decontamination of Surfaces From Extremophile Organisms Using Nonthermal Atmospheric-Pressure Plasmas.* Ieee Transactions on Plasma Science, 2009. **37**(6): p. 866-871.
2. Cooper, M., Y. Yang, G. Fridman, H. Ayan, V.N. Vasilets, A. Gutsol, G. Friedman, and A. Fridman. *Uniform and Filamentary Nature of Continuous-Wave and Pulsed Dielectric Barrier Discharge Plasma*. in *NATO Advanced Study Institute on Plasma Assisted Decontamination of Biological and Chemical Agents*. 2007. Cesme-Izmir, Turkey: Springer.
3. Dobrynin, D., G. Fridman, P. Lelkes, K. Barbee, M. Cooper, H. Ayan, A. Brooks, A. Shereshevsky, M. Balasubramanian, P. Zhu, B. Freedman, C. Kelly, J. Azizkhan-Clifford, S. Kalghatgi, G. Friedman, and A. Fridman. *Mechanisms of Plasma Interaction with Living Tissue*. in *Drexel University Research Day*. 2007. Philadelphia, USA.
4. Morfill, G.E. and A.V. Ivlev, *Complex plasmas: An interdisciplinary research field.* Reviews of Modern Physics, 2009. **81**(4): p. 1353-1404.
5. Fridman, G., M. Peddinghaus, H. Ayan, A. Fridman, M. Balasubramanian, A. Gutsol, A. Brooks, and G. Friedman, *Blood coagulation and living tissue sterilization by floating-electrode dielectric barrier discharge in air.* Plasma Chemistry and Plasma Processing, 2006. **26**(4): p. 425-442.
6. Dobrynin, D., G. Fridman, G. Friedman, and A. Fridman, *Physical and biological mechanisms of direct plasma interaction with living tissue.* New Journal of Physics, 2009(11): p. 115020.
7. Fridman, G., G. Friedman, A. Gutsol, A.B. Shekhter, V.N. Vasilets, and A. Fridman, *Applied Plasma Medicine.* Plasma Processes and Polymers, 2008. **5**(6): p. 503-533.
8. Kalghatgi, S.U., G. Fridman, M. Cooper, G. Nagaraj, M. Peddinghaus, M. Balasubramanian, V.N. Vasilets, A. Gutsol, A. Fridman, and G. Friedman, *Mechanism of Blood Coagulation by Nonthermal Atmospheric Pressure Dielectric Barrier Discharge Plasma.* IEEE Transactions on Plasma Science, 2007. **35**(5, Part 2): p. 1559-1566.
9. D Dobrynin, G.F., G Friedman and A Fridman, *Physical and biological mechanisms of direct plasma interaction with living tissue* New J. Phys., 2009. **11**.
10. Fridman, G., et al., *Blood coagulation and living tissue sterilization by floating-electrode dielectric barrier discharge in air.* Plasma Chemistry and Plasma Processing, 2006. **26**(4): p. 425-442.
11. Fridman, G., A.D. Brooks, M. Balasubramanian, A. Fridman, A. Gutsol, V.N. Vasilets, H. Ayan, and G. Friedman, , *Comparison of Direct and Indirect Effects of Non-Thermal Atmospheric Pressure Plasma on Bacteria.* Plasma Processes and Polymers, 2007. **4**: p. 370-375.
12. Fridman, G., A. Shereshevsky, M. Jost, A. Brooks, A. Fridman, A. Gutsol, V. Vasilets, and G. Friedman, *Floating Electrode Dielectric Barrier Discharge Plasma in Air Promoting Apoptotic Behavior in Melanoma Skin Cancer Cell Lines.* Plasma Chemistry and Plasma Processing, 2007. **27**(2): p. 163-176.
13. Fridman, G., A.B. Shekhter, V.N. Vasilets, G. Friedman, A. Gutsol, and A. Fridman, *Applied Plasma Medicine.* Plasma Processes and Polymers, 2008. **5**(6): p. 503-533.
14. al, Z.-J.C.e., *A realistic brain tissue phantom for intraparenchymal infusion studies.* Journal of neurosurgery, 2004. **101**(2).
15. Ishida, H.K.a.T., *Development of an agar phantom adaptable for simulation of various tissues in the range of 5-40 MHz.* Phys. Med. Biol., 1987. **32**(2): p. 221-226.
16. Dobrynin D, W.A., Kalghatgi S, Park S, Shainsky N, Wasko K, Dumani E, Ownbey R, Joshi S, Sensenig R, Brooks A, *Live Pig Skin Tissue and Wound Toxicity of Cold Plasma Treatment.* Plasma Medicine, 2011. **1**(1): p. 93-108.
17. Gostev, V.a.D., D. *Medical microplasmatron* in *3rd International Workshop on Microplasmas* 2006. Greifswald, Germany.
18. Chakravarthy K, D.D., Fridman G, Friedman G, Murthy S, Fridman A, *Cold Spark Discharge Plasma Treatment of Inflammatory Bowel Disease in an Animal Model of Ulcerative Colitis.* Plasma Medicine, 2011. **1**(1): p. 3-19.
19. S. Kalghatgi, D.D., G. Fridman, M. Cooper, G. Nagaraj, L. Peddinghaus, M. Balasubramanian, K. Barbee, A. Brooks, V. Vasilets, A. Gutsol, A. Fridman, and G. Friedman. **Applications of Non Thermal Atmospheric Pressure Plasma in Medicine**. in *NATO Advanced Study Institute on Plasma Assisted Decontamination of Biological and Chemical Agents*. 2007. Cesme-Izmir, Turkey.

20. Kalghatgi, S.U., G. Fridman, M. Cooper, G. Nagaraj, M. Peddinghaus, M. Balasubramanian, V.N. Vasilets, A. Gutsol, A. Fridman, and G. Friedman, *Mechanism of Blood Coagulation by Nonthermal Atmospheric Pressure Dielectric Barrier Discharge Plasma.* IEEE Transactions on Plasma Science, 2007. **35**(5, Part 2): p. 1559-1566.
21. T Nosenko, T.S.a.G.E.M., *Designing plasmas for chronic wound disinfection.* New J. Phys., 2009. **11**.
22. Stoffels, E., *Cold atmospheric plasma for wound healing: in vitro assessment.*, in *First International Conference on Plasma Medicine (ICPM-1)*2007: Corpus Christi, Texas, USA.
23. Marc Vandamme, E.R., Sebastien Dozias, Julien Sobilo, Stephanie Lerondel, Alain Le Pape, Jean-Michel Pouvesle *Response of Human Glioma U87 Xenografted on Mice to Non Thermal Plasma Treatment*
Plasma Medicine, 2011. **1**(1): p. 27-43.
24. Marc Vandamme, E.R., Sabrina Pesnel, Emerson Barbosa, Sébastien Dozias, Julien Sobilo, Stéphanie Lerondel, Alain Le Pape, Jean-Michel Pouvesle1, *Antitumor Effect of Plasma Treatment on U87 Glioma Xenografts: Preliminary Results.* Plasma Processes and Polymers, 2010. **7**(3-4): p. 264–273.
25. K. Oehmigen, M.H.h., R. Brandenburg, Ch. Wilke, K.-D. Weltmann, and T.v. Woedtke, *The Role of Acidification for Antimicrobial Activity of Atmospheric Pressure Plasma in Liquids.* Plasma Process. Polym., 2010. **7**(3-4): p. 250–257.
26. Danil Dobrynin, G.F., Yurii V. Mukhin, Meghan A. Wynosky-Dolfi, and R.F.R. Judy Rieger, Alexander F. Gutsol, and Alexander Fridman, *Cold Plasma Inactivation of Bacillus cereus and Bacillus anthracis (Anthrax) Spores.* IEEE TPS, 2010. **38**(8): p. 1878-84.
27. Mitsuo Yamamoto, M.N., Masayoshi Sadakata, *Sterilization by $H_2O_2$ droplets under corona discharge.* Journal of Electrostatics, 2002. **56**: p. 173–187.
28. Melly, E., Cowan, A.E., and Setlow, P. , *Studies on the mechanism of killing of Bacillus subtilis spores by hydrogen peroxide.* Journal of Applied Microbiology, 2002. **93**: p. 316-325.
29. M. Rojkind, J.-A.D.-R., N. Nieto and P. Greenwel, *Role of hydrogen peroxide and oxidative stress in healing responses* Cellular and Molecular Life Sciences, 2002. **59**(11): p. 1872-1891.
30. Shanthi Prince, S.M., K. C. Aditya SreeHarsha, Anand Bhandari, Ankit Dua. *An Automated System for Optical Imaging to Characterize Tissue based*
*on Diffuse Reflectance Spectroscopy*. in *Proceedings of the 2nd Biennial IEEE/RAS-EMBS International*
*Conference on Biomedical Robotics and Biomechatronics*. October 19-22, 2008. Scottsdale, AZ, USA.
31. Laura E. Riley, S.A.H., Christopher Henry, Mingui Sun, Robert J. Sclabassi, David Hirsch. *Design of a Phantom Head for the in vitro Testing of Implantable Devices*. in *Bioengineering Conference, 2007. NEBC '07. IEEE 33rd Annual Northeast* 10-11 March 2007 Long Island, NY
32. Markuszewski, H.D.a.R., *Studies on fluorescein—VII: The fluorescence of fluorescein as a function of pH* Talanta, 1989. **36**(3): p. 416-418.
33. Markuszewski, H.D.a.R., *Studies on fluorescein—II: The solubility and acid dissociation constants of fluorescein in water solution.* Talanta, 1985. **32**(2): p. 159-165
34. Diehl, H., *Studies on fluorescein—VIII: Notes on the mathematical work-up of absorbance data for systems with single and multiple dissociations, the limitations of the conventional logarithmic treatment, and the dissociation constants of fluorescein.* Talanta, 1989. **36**(7): p. 799-802
35. Harvey Diehl, N.H.-M., Alta J. Hefley, Linda F. Munson and Richard Markuszewski, *Studies on fluorescein—III : The acid strengths of fluorescein as shown by potentiometric titration.* Talanta, 1986. **33**(11): p. 901-905
36. Lindqvist, M.M.M.a.L., *The pH dependence of fluorescein fluorescence.* Journal of Luminescence, 1975. **10**(6): p. 381-390